# Asynchronous Charge Carrier Injection in Perovskite Light-Emitting Transistors

Maciej Klein,[1,2] Krzysztof Blecharz,[3] Bryan Wei Hao Cheng,[2] Annalisa Bruno[4] and Cesare Soci[1,2*]

[1] *Centre for Disruptive Photonic Technologies, TPI, Nanyang Technological University, 21 Nanyang Link, Singapore 637371*
[2] *Division of Physics and Applied Physics, School of Physical and Mathematical Sciences, Nanyang Technological University, 21 Nanyang Link, Singapore 637371*
[3] *Faculty of Electrical and Control Engineering, Gdansk University of Technology, Narutowicza 11/12, 80-233 Gdansk, Poland*
[4] *Energy Research Institute @ NTU (ERI@N), Nanyang Technological University, 50 Nanyang Drive, Singapore 637553*

*\*Correspondence to: csoci@ntu.edu.sg*

Unbalanced mobility and injection of charge carriers in metal-halide perovskite light-emitting devices pose severe limitations to the efficiency and response time of the electroluminescence. Modulation of gate bias in methylammonium lead iodide light-emitting transistors has proven effective to increase the brightness of light emission, up to MHz frequencies. In this work, we developed a new approach to improve charge carrier injection and enhance electroluminescence of perovskite light-emitting transistors by independent control of drain-source and gate-source bias voltages to compensate for space-charge effects. Optimization of bias pulse synchronization induces a fourfold enhancement of the emission intensity. Interestingly, the optimal phase delay between biasing pulses depends on modulation frequency due to the capacitive nature of the devices, which is well captured by numerical simulations of an equivalent electrical circuit. These results provide new insights into the electroluminescence dynamics of AC-driven perovskite light-emitting transistors and demonstrate an effective strategy to optimize device performance through independent control of amplitude, frequency, and phase of the biasing pulses.

**Keywords**: metal-halide perovskites, light-emitting transistors, pulsed light-emitting devices, electroluminescence modulation, capacitive effects, numerical circuit modeling

# 1. Introduction

In recent years, metal-halide perovskites have shown great potential for optoelectronic devices such as solar cells, photodetectors, X-ray scintillation detectors, light-emitting diodes (LEDs). [1–3] Besides conventional transistors, [4] perovskite light-emitting transistors (PeLETs) integrate two key functionalities of electrical switching and light emission, providing a powerful testbed to study charge transport and recombination processes in semiconducting materials. [5,6] Most recent applications of PeLETs include studies of light-matter interaction in nanophotonic cavities under electrical injection [7] and electrically tunable polarized light sources. [7,8]

Like most hybrid perovskite devices, PeLETs suffer from environmental instability related to intrinsic material limitations, such as temperature-activated trapping, ionic motion, and polarization effects, which reduce their device performance, namely brightness, modulation rate, and uniformity of the recombination zone. Pulsed operation (i.e. AC modulation of the gate bias) has proven to be a viable route to overcome some of these limitations by minimizing ionic vacancy drift and organic cation polarization, and improving space-charge field-assisted injection. [9,10]

Previous studies on AC-operated perovskite light-emitting diodes have focused on the minimization of ionic motion, [11–13] enhancement of brightness and operation stability, [14,15] operating current density reduction, [16] dynamic electroluminescence (EL) response [11,13,17] and direct integration of LEDs into AC power system. [18] This suggests that similar improvements could be achieved in PeLETs through the independent control of source, drain and gate pulse bias parameters.

In this work, we investigate the performance of PeLETs operating in two different pulsed operation modes. Due to the difference in carrier mobilities, injection energy barriers and gate potential screening by mobile charged ions, a number of parameters affect the actual electron and hole charge carrier densities within the recombination zone of the transistor channel. We show that more than fourfold enhancement of electroluminescence intensity can be achieved by controlling the amplitude and electrical polarization of the bias pulses applied to drain and gate terminals and their relative phase, duty cycle, and modulation frequency. This is attributed to the compensation of space-charge effects by alternating currents and to the better overlap of electron and hole distributions within the recombination zone upon asynchronous injection from the electrodes. We develop a comprehensive equivalent circuit model that accounts for both DC and AC response of the PeLET and use LTspice® numerical simulation software to assess the impact of capacitive effects on radiative recombination and overall electroluminescence intensity of the device.

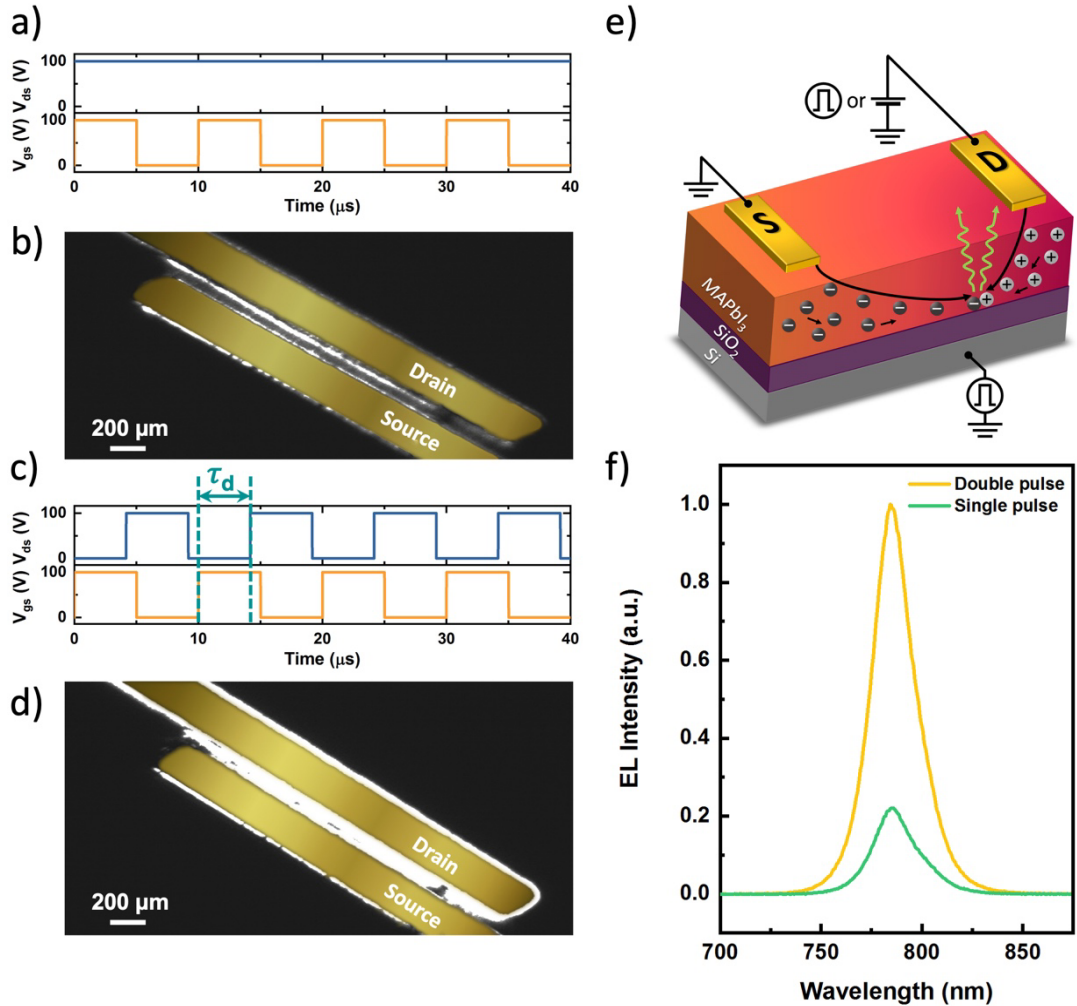

**Figure 1. PeLET emission under single and double pulse mode**. a, c) Waveform charts of biasing voltages ($V_{ds}$ and $V_{gs}$) and b, d) corresponding optical images of the EL emission of PeLET devices under single pulse mode (biasing conditions: $V_{ds}$ =100 V, $V_{gs}$ = 0÷100 V) and double pulse mode (biasing conditions: $V_{ds}$ = 0÷100 V, $V_{gs}$ = 0÷100 V), respectively. The delay time, $\tau_d$, between gate and drain pulses is marked in panel c). e) Schematic of the PeLET device architecture and biasing configuration for i. single pulse mode: DC-biased drain electrode and pulsed-biased gate electrode; ii. double pulse mode: pulsed-biased drain and gate electrodes. f) Comparison of EL spectra under both modes. For double pulse mode phase delay is 150 deg. Measurements were performed at 77 K and 100 kHz modulation frequency. Source and drain electrodes in panels b and d were false colored for better contrast.

## 2. Results and Discussion

The light-emitting transistors based on co-evaporated methylammonium lead iodide ($CH_3NH_3PbI_3$) MAPbI$_3$ perovskite used in this study have similar architecture and electrical characteristics as those described in previous works. [10,19] They show ambipolar charge carrier injection with low-temperature transport dominated by electrons with an on-off ratio of about $10^4$ and field-effect mobility of $(8.6 \pm 0.6) \times 10^{-2}\ cm^2V^{-1}s^{-1}$ at $V_{ds}$ = 60 V. To understand and improve their light emission properties, we investigated the transistor operation in two biasing conditions (**Fig. 1**): i) *Single pulse mode* (Figs. 1a and 1e) where a constant voltage bias is applied to the drain-source (D-S) electrodes, and a pulsed voltage (square wave) is applied to the gate-source (G-S) electrodes; ii) *Double pulse mode* (Figs. 1c and 1e) where a pulsed voltage bias is applied to both drain and gate terminals. In single pulse mode, gate pulses induce the formation of the inversion layer and synchronous injection of charge carriers from source and drain electrodes. As a result of unbalanced carrier mobilities, space-charge effect, and capacitive delay, holes and electrons are not uniformly distributed within the channel. Hence the position of the recombination zone is offset from the center of the channel and the electroluminescence intensity, spatially constrained within a narrow emission line, is weak (Fig. 1b). In double pulse mode, gate and drain pulses are controlled independently. By introducing a phase delay between the pulses, $\Delta\varphi = \frac{\tau_d}{T} \times 360$, where $\tau_d$ is the delay between pulses and T is the modulation period (Fig. 1c), asynchronous injection of charge carriers can be used to achieve spatial and temporal overlap of electron and hole distributions within the recombination zone. The lateral electric field between source and drain electrodes induces charge carriers to drift and recombine throughout the channel, resulting in a widely spread emission zone (Fig. 1d). At the optimal phase delay of $\Delta\varphi$ = 150 deg between biasing pulses, and modulation frequency of 100 kHz, the integrated EL intensity in double pulse mode is 4.5 times higher than in single pulse mode (Fig. 1f).

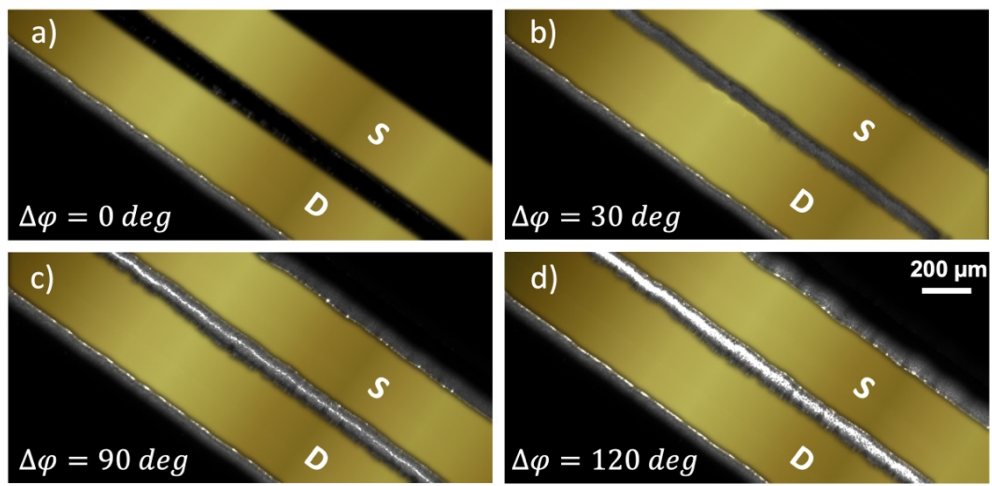

**Figure 2. Electroluminescence of PeLET under asynchronous injection.** a-d) optical images of the EL emission of PeLET devices at various phase delays between drain and gate

pulses. Dephasing the source relative to the gate pulses from Δφ = 0 deg (a) to Δφ = 120 deg (d) modifies the recombination zone within the channel and progressively increases the brightness of EL. Measurements were performed at 77 K and 500 kHz modulation frequency. Source and drain electrodes were false colored for better contrast.

The electroluminescence observed in the AC-driven PeLETs appears to originate from two separate mechanisms. EL emitted from within the transistor channel is attributed to band-to-band radiative recombination of electrons and holes injected from the top S and D electrodes, while EL emitted from beneath and in close proximity of the drain and source electrodes is associated with AC field-induced space-charge recombination. [9,20] Interestingly, these two effects can be spatially resolved in the optical images of the PeLET operated in the double pulse mode with different phase delays (and the difference becomes more apparent at high modulation frequency, e.g. 500 kHz), **Fig. 2**. At phase delay of Δφ = 0 deg (Fig. 2a), AC field-induced recombination around the top electrodes is dominant while at Δφ = 30 deg (Fig. 2b) light is emitted uniformly throughout the channel, suggesting that both recombination processes may be equally involved. Further increase of the phase delay leads to strong EL emission from the center of the channel governed by band-to-band recombination of the injected charge carriers (Fig. 2c), which saturates at Δφ = 120 deg (Fig. 2d). Continuous-frame video showing the tunability of the EL intensity and the position of the emission zone as a function of Δφ is provided as Supplementary Movie 1 in Supporting Information.

In addition to the relative phase of the drain and gate pulses, double pulse mode operation provides additional degrees of freedom to control the EL intensity of PeLETs, i.e. duty cycle ($\delta = \frac{\tau_{on}}{T} \times 100\%$, where $\tau_{on}$ is pulse width) and modulation frequency ($f$) of the individual drain and gate pulses (**Fig. 3**). The dependence of the integrated EL intensity on the phase between the drain and gate pulses (Δφ), at a constant duty cycle of δ = 50% and modulation frequency of 100 kHz (pulse width of $\tau_{on}$ = 5 μs), is shown in Fig. 3a (a diagram of the corresponding biasing pulse trains is shown in Fig. S1). Surprisingly, the EL intensity emitted by the PeLET does not reflect the temporal overlap between the drain and gate pulses as a function of their dephasing. In fact, the maximum emission intensity occurs when the gate and drain pulses are out of phase and their temporal overlap is minimal (Δφ=150-180 deg). Drain and gate duty cycles have a similar effect on the EL intensity, with maxima occurring at δ = 50% (Figs. 3b and 3c). Thus, neither prolonging the formation of the inversion layer by increasing the gate duty cycle nor prolonging charge carrier injection by increasing the drain duty cycle increases the emission intensity as one may expect. This indicates that the charge carrier recombination dynamics responsible for light emission occurs on a faster time scale than the duration of the biasing pulses. Note that the second maximum observed at δ = 70% (Fig. 3b) occurs when the drain pulse overlaps with the gate pulse of the following cycle (Fig. S2a). Furthermore, the gate duty cycle yielding maximum EL intensity (Fig. 3c) corresponds to minimal overlap between gate and drain pulses (Fig. S2b). At a larger modulation frequency of 500 kHz ($\tau_{on}$ = 1 μs), the highest EL intensity is obtained at a smaller phase delay of 120 deg, Fig. S3a (corresponding to a pulse delay of 0.67 μs, Fig. S1b), while the optimal gate and drain duty cycles remain 50% (Figs. S3b, S3c and S4), consistent with the results at 100 kHz.

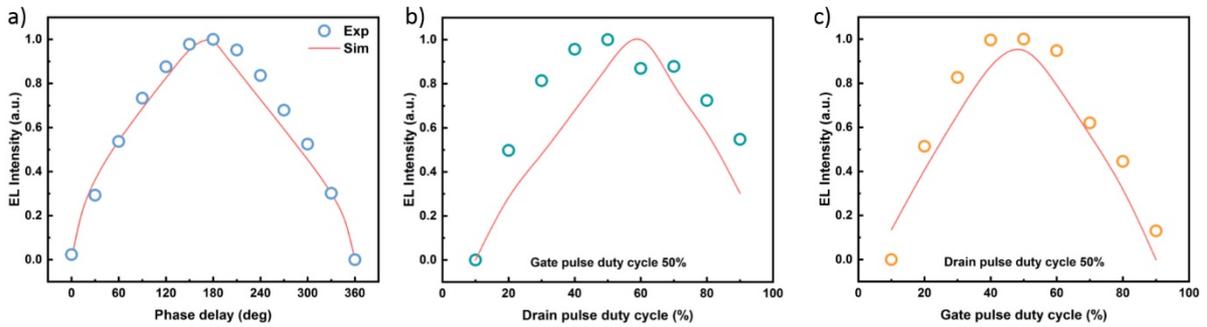

**Figure 3. Electroluminescence dynamics versus the pulse duty cycle.** a) Phase delay dependent EL intensity and EL intensity dynamics versus b) drain pulse duty cycle (constant gate pulse δ = 50%) and c) gate pulse duty cycle (constant drain pulse δ = 50%). Measurement conditions: 100 kHz modulation frequency, 150 deg phase delay (for panels b and c).

The optimal phase delay between pulses may be related to the formation time of the inversion layer within the perovskite film, with its peculiar dependence on modulation frequency reflecting the inherent capacitive effects of the perovskite. A simple model to describe the operation of FETs consists of two separate metal-insulator-semiconductor (MIS) diodes representing the gate/source and the gate/drain electrode pairs, with the intrinsic capacitance of the diodes defining the dynamic response of the transistor. To apply this model to our devices, the capacitance $C(f)$ of an equivalent MIS structure, Si/SiO$_2$ (500 nm)/MAPbI$_3$ (400 nm)/Au (100 nm), was measured by impedance spectroscopy as a function of modulation frequency (Fig. S5). In such MIS structure, the total capacitance $C$ is given by the combination of the insulator capacitance $C_i$ (SiO$_2$) and the semiconductor layer capacitance $C_s$ (MAPbI$_3$) connected in series: $C = \frac{C_i C_s}{C_i + C_s}$.[21] A significant decrease in capacitance at frequencies above 7×10$^4$ Hz is clearly visible in the data. Such frequency dependence of $C(f)$ is consistent with prior literature reports for spin-coated perovskite films, that attributed the reduction of $C_s$ at high frequencies to polarization effects.[22,23] The overall capacitance is related to the time constant of the equivalent RC circuit, τ = RC, where R is the equivalent circuit resistance. Hence, a decrease of $C(f)$ at a high modulation frequency shortens τ.

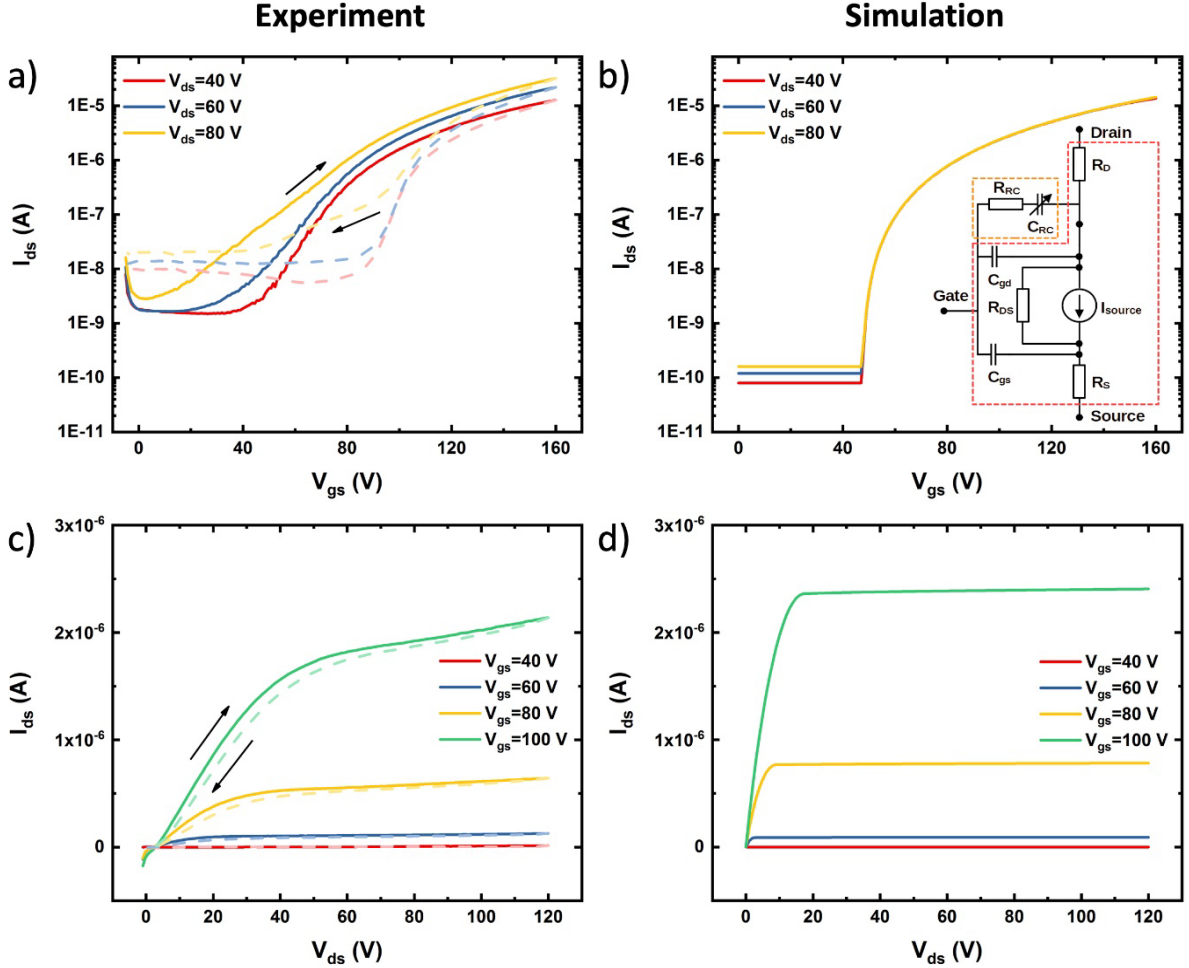

**Figure 4. DC electrical characteristics of the PeLETs.** a,c) Measured and b,d) simulated *n*-type a,b) transfer and c,d) output characteristics under the constant drain ($V_{ds}$) and gate ($V_{gs}$) bias indicated in the panels, at 77 K. Solid lines were obtained in the forward direction while dashed lines in the reverse direction of the voltage sweep. The inset of panel b) shows equivalent circuit of the developed PeLET model.

DC electrical characteristics of the PeLETs are shown in **Fig. 4**. The electrical model was implemented in LTspice® software using the mathematical Grove-Frohman model for metal-oxide-semiconductor field-effect transistor (MOSFET) and the experimentally determined capacitances, as shown in the red dashed box in the inset of Fig. 4b. [24] The measured *n*-type transfer and output characteristics of PeLETs shown in Figs. 4a,c are in good agreement with the simulated curves (Figs. 4b,d). The gain of $I_{ds}$ with increasing $V_{gs}$ voltage in transconductance curves as well as $I_{ds}$ saturation currents are comparable. The main discrepancy between simulations and experiments is the lack of electrical hysteresis arising from slow ionic motion and organic cation polarization disorder. These effects are not included in the model since they are effectively reduced upon AC-modulation at frequencies >$10^4$ Hz.[9,10,25,26]

The complex dependence of the EL intensity signal on the phase delay observed in Fig. 3 cannot be solely explained by the simple model of metal-oxide-semiconductor field-effect transistor (MOSFET). To unveil the nature of the transport dynamics in the PeLET, we performed transient electroluminescence measurements under double pulse mode operation and implemented a modified version of the MOSFET equivalent circuit (**Fig. 5**). The new electrical circuit model includes an additional RC branch, comprising $R_{RC}$ and $C_{RC}$, driven by the gate voltage, as shown in the orange dashed box in the inset of Fig. 4b. The generalized rate equation for the charge population dynamics upon electrical injection within the width $d$ of the PeLET recombination zone can be written as $\frac{dn}{dt} = \frac{j}{qd} - k_r n$, where $n(t)$ is the density of charge carriers, $j(t)$ is the current density, $q$ is the elementary charge, and $k_r$ is the recombination rate.[27] At steady state, i.e., $\frac{dn}{dt} = 0$, the rate equation gives $n(t) = \frac{j(t)}{qdk_r}$. Assuming that the electroluminescence emission intensity in PeLETs is proportional to the drain current flowing in the transistor channel, $I_{ds}$, similar to what is typically done to simulate emission characteristics of LEDs, the time dependence of the PeLET electroluminescence upon current injection in double pulse mode (Fig. 5a) can be directly compared with the time evolution of the carrier density (Fig. 5b).

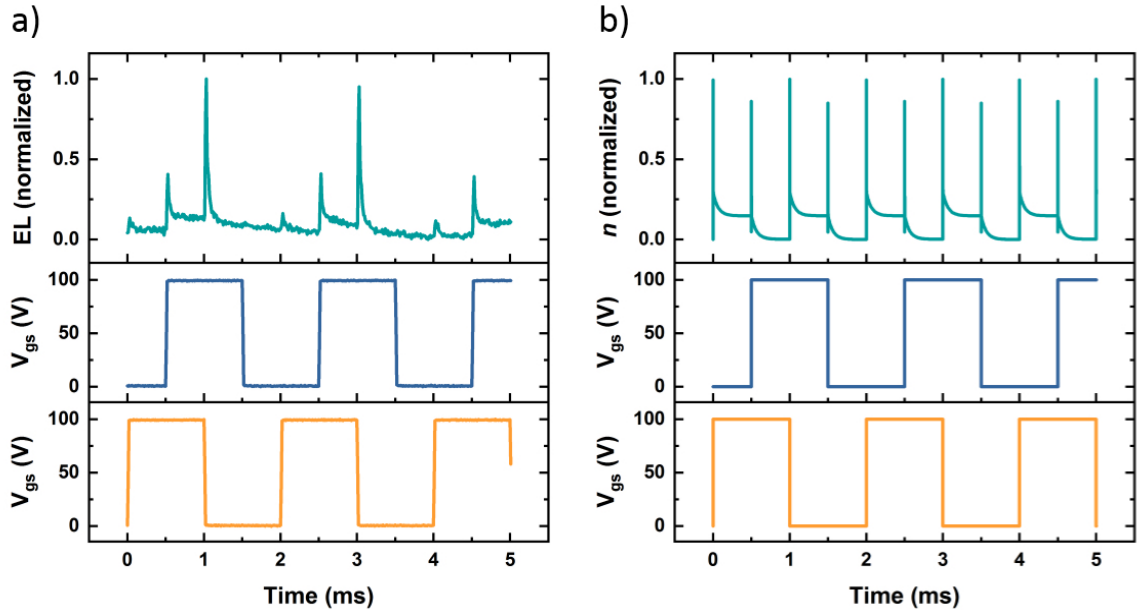

**Figure 5. Transient electroluminescence of the PeLETs.** Time dependence of a) measured electroluminescence and b) simulated carrier density. Schematic diagrams of the biasing pulse

trains are shown in the bottom panels. Measurement conditions: 500 Hz modulation frequency, 90 deg phase delay.

The recorded EL signal shows sharp light emission peaks of lower intensity at the beginning of D-S pulses and high-intensity peaks at the end of G-S pulses (Fig. 5a). This may be attributed to the capacitive charging and discharging of the device, wherein the charge carriers are accumulated in the perovskite film, at the interface with the injecting electrodes, as previously observed for the EL dynamics of perovskite solar cells. [11] This suggests that under asynchronous pulsed conditions, source/drain and gate electrodes are capacitively coupled and the PeLET acts like a field-induced capacitive light-emitting device.[28] The EL decay is multiexponential, with three characteristic decay times, $\tau_1 = 7.1 \times 10^{-7}$ s, $\tau_2 = 2.7 \times 10^{-5}$ s, and $\tau_3 = 2.4 \times 10^{-4}$ s. The two fastest components may be attributed to the radiative decay responsible for light emission (the $\tau_1$ time is not fully resolved due to the limitations of our experimental setup), while the sub-millisecond component could be ascribed to the slower screening effects of ionic motion and polarization. [9,25] Similarly, the simulated carrier density shows emission peaks and multiexponential decay dynamics that are in fairly good agreement with the experiments (Fig. 5b). Here, the fast component of the simulated decay, $\tau_1 = 2.4 \times 10^{-7}$ s, is defined by the MOSFET part of the model with the $C_{gs}$, and $C_{gd}$ capacitances set to the experimental values, while the two slower components are induced by the RC series circuit with $R_{RC}C_{RC}$ time constants $\tau_2 = 3.0 \times 10^{-5}$ and $\tau_3 = 1.0 \times 10^{-4}$ s. $\tau_2$ arises from the linear component of $C_{RC}$, while $\tau_3$ stems from the nonlinearity of $C_{RC}$ introduced by an additional parallel resistance and a parallel capacitance, which accounts for energy dissipation in the circuit.

Further validation of the electrical circuit model with the determined time constants is provided by the dependence of the carrier density on dephasing and duty cycle of the biasing pulses at a given modulation frequency (red lines in Fig. 3 and S3). The good agreement between the integrated carrier density and measured EL intensity over the entire range of biasing conditions, with no adjustable parameters, confirms the reliability of the model to describe the complex transport characteristics of PeLETs. Such model provides valuable insights on the interplay between processes occurring at broadly different time scales, namely intrinsic radiative recombination of the perovskite excited states in the sub-microsecond regime, extrinsic capacitive effects related to device architecture at the microsecond time scale, as well as energy dissipation induced by ionic screening and polarization effects that persist up to milliseconds.

## 3. Conclusions

In summary, we demonstrated that asynchronous charge injection yields a significant increase of the electroluminescence intensity of AC-driven perovskite light-emitting transistors. This is attributed to the compensation of ionic and polarization effects and efficient utilization of injected charge carriers and accumulated space-charges upon the formation of an inversion layer. At a given modulation frequency, the optimal phase delay between the gate and drain-source pulses is determined by the capacitive properties of the perovskite layer. Based on time-resolved and phase delay dependent electroluminescence characteristics of the PeLETs operated in the double pulse mode, we developed a comprehensive equivalent circuit model that accounts for both DC and AC response, and unveils the impact of capacitive effects on radiative recombination and overall electroluminescence intensity. The reliability of the model was validated by numerical simulations of the PeLET electrical characteristics and of the carrier density dependence on dephasing and duty cycle of the biasing pulses. The understanding of the relevant processes and their characteristic response time in PeLET devices provides new ways to optimize their architecture and output characteristics, potentially yielding orders of magnitude improvement in electroluminescence brightness and efficiency with optimized device structure and asynchronous biasing conditions. We foresee that the availability of bright and fast switchable PeLETs operating in room temperature, in conjunction with advanced metaoptics concepts recently demonstrated on this platform, [7,8] will substantially advance their application in lighting, active matrix displays, and optical wireless communication. [29–31]

## 4. Experimental Section

*Device fabrication and electrical characterization*

PeLETs were fabricated in a bottom-gate and top-contact configuration on heavily *p*-doped Si substrates with 500 nm thermally grown $SiO_2$ (capacitance of 6.9 nFcm$^{-2}$) layer, following the processes described in previous work. [10] $MAPbI_3$ films were deposited by thermal co-evaporation of $PbI_2$ powder (TCI) and methylammonium iodide (MAI) powder (Lumtec) from effusion sources in high vacuum (pressure $< 1 \times 10^{-5}\ mbar$) using the conditions described in reference 19. The top contact gold source and drain electrodes (channel length: 100 μm, channel width: 1 mm) were deposited by thermal evaporation through a shadow mask. Electrical measurements of the transistors were carried out at 77 K in the dark and under the vacuum (10$^{-3}$ mbar) using a temperature-controlled electrical probing stage (Linkam HFS600E-PB4/PB2). The electrical characteristics were acquired with a 2-channel precision source/measure unit (Agilent B2902A). Charge-carrier mobilities were extracted from the forward sweeping of transfer characteristics obtained at $V_{ds} = \pm 60$ V, using the conventional equation for metal-oxide semiconductor (MOS) transistors in the saturation regime: $\mu_{sat} = \frac{2L}{WC_i}\left(\frac{\partial\sqrt{I_{ds}}}{\partial V_{gs}}\right)^2$.

*Electroluminescence measurements*

Pulsed electroluminescence measurements were performed applying a square wave bias to the PeLET gate and drain electrodes, using a 2-channel arbitrary waveform generator (Rigol DG832) that allows for precise signals phase adjustment. The two output waveforms were subsequently amplified by separate high-voltage amplifiers (Falco Systems WMA-300), with rise time and fall time < 50 ns. Optical images were acquired by a sCMOS cooled scientific camera (PCO Edge 3.1m) coupled to an optical microscope (Motic PSM-1000). The electroluminescence spectra were collected using a fiber-coupled spectrometer (Avantes AvaSpec ULS-RS-TEC). The EL intensity was obtained by the integration of the measured spectrum for a given phase delay. Transient electroluminescence response of the PeLETs was collected by using a Si photodiode (Newport 818-UV) connected to a low-noise current preamplifier (Stanford Research Systems SR570) and a digital oscilloscope (LeCroy WaveSurfer 104MXs-B). A trigger for the oscilloscope was taken from the waveform generator. The time constant of the amplifier is ≈ 0.35 μs. All measurements were performed at a low temperature (77 K) to minimize the effects of ionic drift and maximize electroluminescence intensity. Source and drain electrodes in Figs. 1b,d and 2 were false-colored using GIMP software.

*Impedance spectroscopy*

Impedance spectroscopy measurements were conducted using a vertical multilayer device consisting of Si/SiO$_2$ (500 nm)/MAPbI$_3$ (400 nm)/Au (100 nm). The perovskite layer and the Au top electrode were deposited in the same manner as for PeFET transistors. Impedance measurements were performed using a potentiostat (BioLogic SP-200) over a frequency range of 100 Hz to 1 MHz, with an applied AC voltage of 30 mV. The measurements were conducted in the dark to avoid photocapacitive effects. [32]

*Simulations*

Numerical simulations were performed in LTspice® XVII (Analog Devices, Inc.) software based on the standard mathematical Grove-Frohman MOSFET model (MOSFET Model LEVEL 2).[24] The modified trap integration method was applied, while the maximum timestep was 3 orders of magnitude lower than the voltage modulation period. Parameters such as field-effect mobility, $C_{gs}$, and $C_{gd}$ capacitances as well as device dimensions used in the simulations were defined based on the experiments.

**Acknowledgments**

The authors would like to acknowledge Miloš Petrović for the fruitful discussions on time-resolved measurements and Li Jia for providing the co-evaporated perovskite films used in these studies. Research was supported by the A*STAR-AME programmatic fund on Nanoantenna Spatial Light Modulators for Next-Gen Display Technologies (Grant no. A18A7b0058) and the Singapore Ministry of Education MOE Tier 3 (Grant no. MOE2016-T3-1-006). K.B. acknowledges support from the IDUB Ventus-Hydrogenii Gdansk Tech Program (Grant no. DEC-3/2022/IDUB/VHR).

**Data availability**

The authors declare that all data supporting the findings of this study are available within this article and its supplementary information and are openly available in NTU research data repository DR-NTU (Data) at https://doi.org/10.21979/N9/FPXQPA. Additional data related to this paper may be requested from the authors.

**Author contributions**

M.K. and C.S. conceived the idea. M.K. fabricated the light-emitting devices and carried out all electrical and optical measurements with the support of B.W.H.C.. A.B. was responsible for perovskite films preparation. M.K. and K.B. developed the equivalent circuit model and performed numerical simulations. M.K. and C.S. performed data analysis and wrote the manuscript with inputs from all authors. C.S. supervised the work.

**References**

[1] L. Chouhan, S. Ghimire, C. Subrahmanyam, T. Miyasaka, V. Biju, *Chemical Society reviews* **2020**, *49*, 2869.
[2] S. D. Stranks, H. J. Snaith, *Nature Nanotechnology* **2015**, *10*, 391.
[3] M. D. Birowosuto, D. Cortecchia, W. Drozdowski, K. Brylew, W. Lachmanski, A. Bruno, C. Soci, *Scientific Reports* **2016**, *6*, 37254.
[4] C. R. Kagan, D. B. Mitzi, C. D. Dimitrakopoulos, *Science* **1999**, *286*, 945.
[5] X. Y. Chin, D. Cortecchia, J. Yin, A. Bruno, C. Soci, *Nature Communications* **2015**, *6*, 7383.
[6] J. Zaumseil, *Advanced Functional Materials* **2020**, *30*, 1905269.
[7] M. Klein, Y. Wang, J. Tian, S. T. Ha, R. Paniagua-Domínguez, A. I. Kuznetsov, G. Adamo, C. Soci, *Advanced Materials* **2022**, 2207317.
[8] Y. Wang, J. Tian, M. Klein, G. Adamo, S. T. Ha, C. Soci, *ArXiv* **2022**, 2209.05810.
[9] F. Maddalena, X. Y. Chin, D. Cortecchia, A. Bruno, C. Soci, *ACS Applied Materials and Interfaces* **2018**, *10*, 37316.
[10] M. Klein, J. Li, A. Bruno, C. Soci, *Advanced Electronic Materials* **2021**, *7*, 2100403.
[11] R. Gegevičius, M. Franckevičius, J. Chmeliov, W. Tress, V. Gulbinas, *Journal of Physical Chemistry Letters* **2019**, *10*, 1779.
[12] H. Kim, L. Zhao, J. S. Price, A. J. Grede, K. Roh, A. N. Brigeman, M. Lopez, B. P. Rand, N. C. Giebink, *Nature Communications* **2018**, *9*, 4893.
[13] N. K. Kumawat, W. Tress, F. Gao, *Nature Communications* **2021**, *12*, 4899.
[14] J. Liu, X. Sheng, Y. Wu, D. Li, J. Bao, Y. Ji, Z. Lin, X. Xu, L. Yu, J. Xu, K. Chen, *Advanced Optical Materials* **2018**, *6*, 1700897.
[15] X. Cheng, Z. Zang, K. Yuan, T. Wang, K. Watanabe, T. Taniguchi, L. Dai, Y. Ye, *Applied Physics Letters* **2020**, *116*, 263103.
[16] J. Liu, Z. Lu, X. Zhang, Y. Zhang, H. Ma, Y. Ji, X. Xu, L. Yu, J. Xu, K. Chen, *Nanomaterials* **2018**, *8*, 974.
[17] R. Chakraborty, G. Paul, A. J. Pal, *Physical Review Applied* **2020**, *14*, 024006.
[18] X. Liu, D. Yu, C. Huo, X. Song, Y. Gao, S. Zhang, H. Zeng, *Advanced Optical Materials* **2018**, *6*, 1800206.
[19] J. Li, H. Wang, X. Y. Chin, H. A. Dewi, K. Vergeer, T. W. Goh, J. W. M. Lim, J. H. Lew, K. P. Loh, C. Soci, T. C. Sum, H. J. Bolink, N. Mathews, S. Mhaisalkar, A. Bruno, *Joule* **2020**, *4*, 1035.
[20] X. Liu, J. Kjelstrup-Hansen, H. Boudinov, H. G. Rubahn, *Organic Electronics* **2011**, *12*, 1724.
[21] S. M. Sze, K. K. Ng, *Physics of Semiconductor Devices*, 3rd ed., John Wiley & Sons, Ltd., Hoboken, NJ, **2007**.
[22] S. P. Senanayak, B. Yang, T. H. Thomas, N. Giesbrecht, W. Huang, E. Gann, B. Nair, K. Goedel, S. Guha, X. Moya, C. R. McNeill, P. Docampo, A. Sadhanala, R. H. Friend, H. Sirringhaus, *Science Advances* **2017**, *3*, e1601935.
[23] A. Bruno, D. Cortecchia, X. Y. Chin, K. Fu, P. P. Boix, S. Mhaisalkar, C. Soci, *Advanced Energy Materials* **2017**, *7*, 1700265.
[24] A. Vladimirescu, S. Liu, *The Simulation of MOS Integrated Circuits Using SPICE2*, Report No. UCB/ERL M80/7, **1980**.
[25] A. M. A. Leguy, J. M. Frost, A. P. McMahon, V. G. Sakai, W. Kochelmann, C. Law, X. Li, F. Foglia, A. Walsh, B. C. O'Regan, J. Nelson, J. T. Cabral, P. R. F. Barnes, *Nature Communications* **2015**, *6*, 7124.
[26] S. P. Senanayak, B. Yang, T. H. Thomas, N. Giesbrecht, W. Huang, E. Gann, B. Nair, K. Goedel, S. Guha, X. Moya, C. R. McNeill, P. Docampo, A. Sadhanala, R. H. Friend, H. Sirringhaus, *Science Advances* **2017**, *3*, e1601935.

[27] S. R. Forrest, *Organic electronics: Foundations to applications*, Oxford University Press, **2020**.
[28] Y. Chen, Y. Xia, G. M. Smith, D. L. Carroll, *Advanced Materials* **2014**, *26*, 8133.
[29] S. Kahmann, A. Shulga, M. A. Loi, *Adv Funct Mater* **2020**, *30*, 1904174.
[30] H. Elgala, R. Mesleh, H. Haas, *IEEE Communications Magazine* **2011**, *49*, 56.
[31] L. Zhao, K. Roh, S. Kacmoli, K. Al Kurdi, X. Liu, S. Barlow, S. R. Marder, C. Gmachl, B. P. Rand, *Advanced Materials* **2021**, *33*, 2104867.
[32] E. J. Juarez-Perez, R. S. Sanchez, L. Badia, G. Garcia-Belmonte, Y. S. Kang, I. Mora-Sero, J. Bisquert, *Journal of Physical Chemistry Letters* **2014**, *5*, 2390.

Supporting Information for

# Asynchronous Charge Carrier Injection in Perovskite Light-Emitting Transistors

Maciej Klein,[1,2] Krzysztof Blecharz,[3] Bryan Wei Hao Cheng,[2]

Annalisa Bruno[4] and Cesare Soci[1,2]*

[1] *Centre for Disruptive Photonic Technologies, TPI, Nanyang Technological University, 21 Nanyang Link, Singapore 637371*
[2] *Division of Physics and Applied Physics, School of Physical and Mathematical Sciences, Nanyang Technological University, 21 Nanyang Link, Singapore 637371*
[3] *Faculty of Electrical and Control Engineering, Gdansk University of Technology, Narutowicza 11/12, 80-233 Gdansk, Poland*
[4] *Energy Research Institute @ NTU (ERI@N), Nanyang Technological University, 50 Nanyang Drive, Singapore 637553*

*\*Correspondence to: [csoci@ntu.edu.sg](mailto:csoci@ntu.edu.sg)*

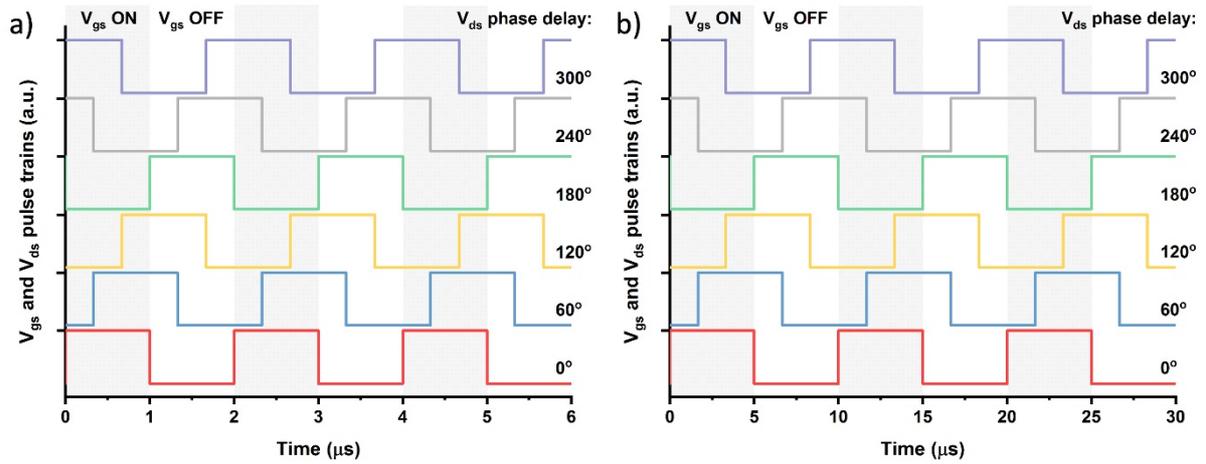

**Figure S1.** Schematic diagram of biasing pulse trains as a function of the phase delay for a) 100 kHz and b) 500 kHz modulation frequency, wherein the shaded area represents gate pulse train. Drain pulses at corresponding phase delays are indicated in the panels.

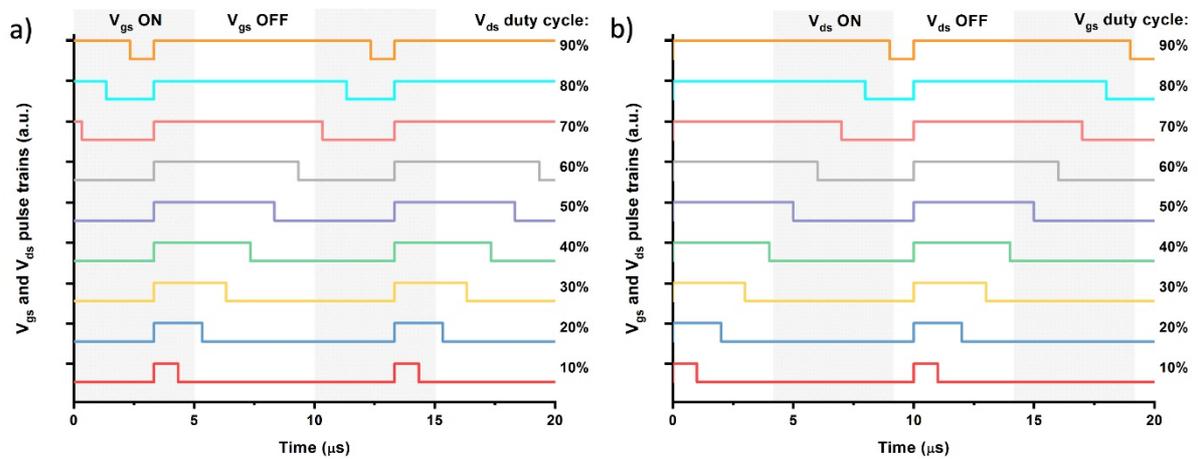

**Figure S2.** Schematic diagram of biasing pulse trains as a function of a) drain pulse duty cycle (for constant gate pulse δ = 50% represented by the shaded area) and b) gate pulse duty cycle (for constant drain pulse δ = 50% represented by the shaded area) under the following conditions: 100 kHz modulation frequency and 150 deg phase delay.

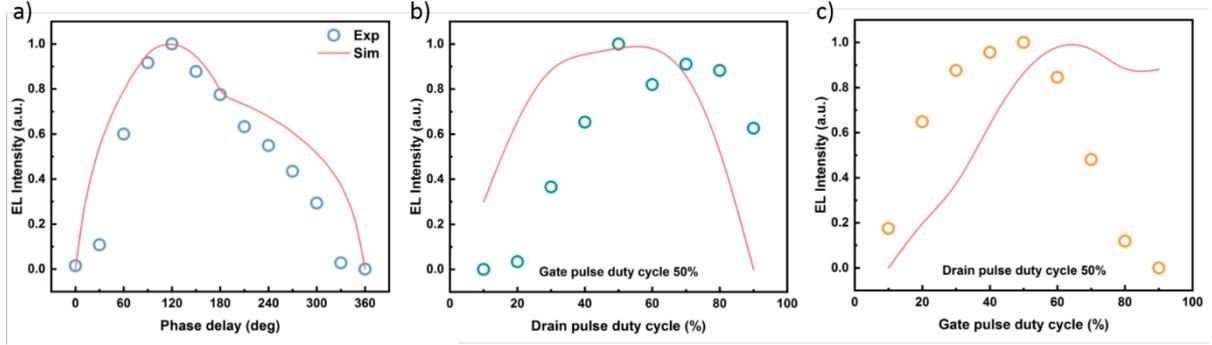

**Figure S3.** a) Phase delay dependent EL intensity and EL intensity dynamics versus b) drain pulse duty cycle (gate pulse δ = 50%) and c) gate pulse duty cycle (drain pulse δ = 50%). Measurement conditions: 500 kHz modulation frequency, 120 deg phase delay (for panels b and c).

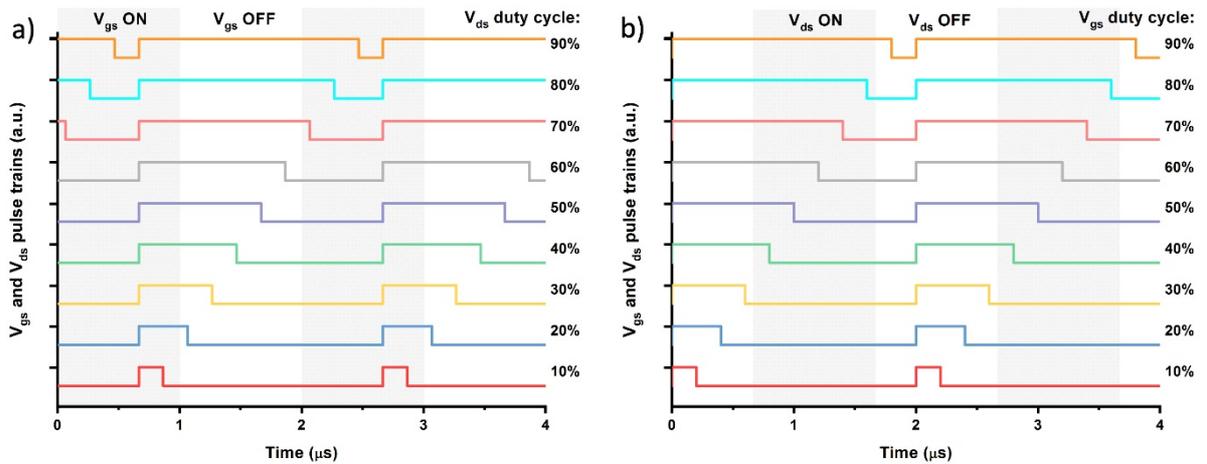

**Figure S4.** Schematic diagram of biasing pulse trains as a function of a) drain pulse duty cycle (for constant gate δ = 50% represented by the shaded area) and b) gate pulse duty cycle (for constant drain δ = 50% represented by the shaded area) under the following conditions: 500 kHz modulation frequency and 120 deg phase delay.

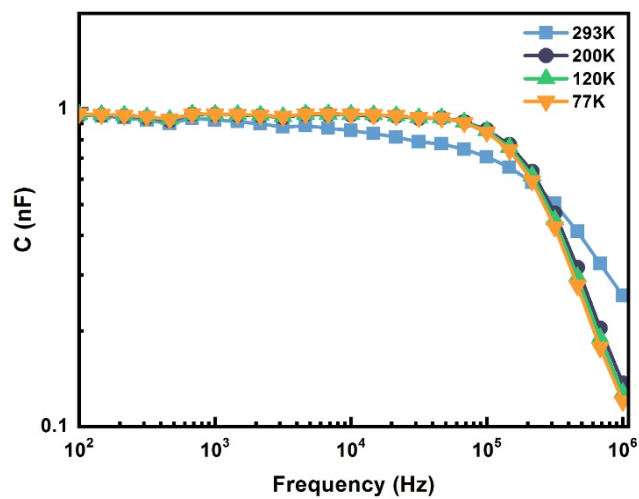

**Figure S5.** The total capacitance of a sandwich device: Si/SiO$_2$ (500 nm)/ MAPbI$_3$ (400 nm)/Au (100 nm), as a function of temperature, with an applied AC voltage of 30 mV, measured at 77 K.

**Supplementary Movie 1.** Light emission from the PeLET under double pulse mode at 77 K and following biasing conditions: V$_{ds}$ = 0÷100 V, V$_{gs}$ = 0÷100 V, 500 kHz modulation frequency. Video recorded while sweeping the dephasing of the source relative to the gate pulses Δφ from 0 to 360 deg.